\newcommand{\ba}{\begin{array}}
\newcommand{\ea}{\end{array}}
\newcommand{\bean}{\begin{eqnarray*}}
\newcommand{\eean}{\end{eqnarray*}}

\newcommand{\bmx}[1]{\left(\begin{array}{*{#1}{c}}}
\newcommand{\emx}{\end{array}\right)}
\newcommand{\bmxw}[1]{\renewcommand{\arraystretch}{2}\left(\begin{array}{*{#1}{c}}}
\newcommand{\bmxww}[1]{\renewcommand{\arraystretch}{2.5}\left(\begin{array}{*{#1}{c}}}
\newcommand{\bdet}[1]{\renewcommand{\arraystretch}{1.2}
	\left|\begin{array}{*{#1}{c}}}
\newcommand{\edet}{\end{array}\right|\renewcommand{\arraystretch}{1}}
\newcommand{\beq}{\begin{equation}}
\newcommand{\eeq}{\end{equation}}
\newcommand{\bea}{\begin{eqnarray}}
\newcommand{\eea}{\end{eqnarray}}

\newcommand{\ditem}[1]{\item[$\diamond$]}

\newcommand{\bit}{\begin{itemize}}
\newcommand{\eit}{\end{itemize}}

\documentclass[aps,prl,twocolumn,groupedaddress]{revtex4}
\usepackage{subfigure}
\usepackage{epsfig}
\usepackage{amssymb}

\begin{document}


\title{Numerical Studies of Quantum Oscillations in the Superconducting vortex mixed state}


\author{Kuang-Ting Chen}
\email[]{timchen@mit.edu}
\affiliation{Center for Theoretical Physics, Massachusetts Institute of Technology, Cambridge, MA 02139}
\author{Patrick A. Lee}
\affiliation{Department of Physics, Massachusetts Institute of Technology, Cambridge, MA 02139}


\date{\today}

\begin{abstract}
We studied the quantum oscillations in superconducting vortex mixed states with d-wave pairing. We showed that the Onsager relation does not always apply. Furthermore, at mean field level the quantum oscillations from electron pockets are suppressed by the pairing. We conclude that an interpretation of the experimental results asccoming  from the four hole pockets created by a $(\pi,\pi)$ folding cannot be ruled out.
\end{abstract}


\maketitle

Recent experiments have shown quantum oscillations in underdoped YBCO samples in strong magnetic field$\sim$45 T \cite{Doiron07,Bangura08,Yelland08,Jaudet08,Sebastian08}. This observation has been interpreted as a signal of the underlying Fermi surface (FS). The period of the oscillation is large, which corresponds to a very small FS, and implies that the FS has been reconstructed, probably by some translational symmetry breaking order in the ground state \cite{Chakravarty08,Millis07,Jdimov,podolsky}. However, for both materials that have been studied, the simplest construction, a $(\pi,\pi)$ folding forming four hole pockets in the original Brillouin Zone (BZ), would imply a pocket too small comparing to the nominal doping by about $25\%$ \cite{Doiron07,Yelland08}. This partly motivated a number of workers to intepret the data in terms of more complicated reconstruction, such as incommensurate spin density wave (SDW) \cite{Sebastian08,Millis07}. Furthermore, the measured Hall effect is negative \cite{LeBoeuf07} and this has led LeBoeuf et al to propose that the quantum oscillations originate from electron pockets. Whether the negative Hall effect is due to flux flow \cite{Ong} is currently a subject under debate. 

We believe that in these experiments the samples are still in a vortex mixed state. One evidence is the measured torque hysteresis \cite{Sebastian08}, which implies that vortices exist at least up to 45 T. Furthermore, the commonly quoted core size of 20$\,A^\circ$ (see for example, extrapolation based on STM measurements \cite{shpan} as well as Nernst measurements \cite{luli}) lead to an estimate of $H_{c2}$ of 100 T. Therefore, in order to interpret the oscillations data one has to understand the quantum oscillations in the mixed state. 

Up to now all discussions assume that quantum oscillations in the mixed state maintain the same frequency as in the normal state. This is true of all experiments performed up to date where it is possible to scan the magnetic field across $H_{c2}$ \cite{tjanssen,wosnitze}. However, we note that all previous experiments have been done on conventional s-wave low $T_c$ superconductors, with the possible exception of the organics which may be d-wave, and the high $T_c$ cuprates may be in quite a different parameter regime. For example, the coherence length $\xi_0$ is very short, of order 4 or 5 lattice constants and is the consequence of a large energy gap $\Delta_0$. The number of Landau levels $N$ is about 10 in the high $T_c$ experiments, as opposed to 100's or 1000's for coventional superconductors. This means that the semiclassical orbit encloses 10 flux quanta, i.e., 20 vortex cores. There are good reasons to believe that the pairing amplitude and the gap scale $\Delta_0$ are very robust in the underdoped cuprates. In contrast, the commonly used theory of Maki \cite{Maki} and Stephen \cite{Stephen} considers the lowest order scattering of the normal state quasiparticles by a pairing order parameter which is random with short range correlations. It is not at all clear that this picture applies to the current problem. Another parameter which clearly distinguishes the cuprate from other superconductors is $k_F\xi_0$ which is less than or of order unity if $k_F$ is the size of the observed pocket while $k_F\xi_0>>1$ for conventional superconductors.

In this paper we address the question of whether the traditional picture continues to hold in a parameter regime which has not been tested experimentally. We take as our model the Bogoliubov-de Gennes (BdG) equations with a variety of vortex coordinates and competing order parameters, which can take on arbitrary spatial dependence. We take the coherence length and the pair field as parameters and make no attempt to solve the problem self-consistently. Since we do not expect the BdG equations to be the correct microscopic theory for the underdoped cuprates, it makes little sense to determine these parameters self-consistently. Rather, we treat this as a phenomenological model. To the extent that the proximity to the Mott transition is not captured by by this phenomenology, the application of our theory to high $T_c$ problems should be treated with caution. With these caveates, we write down a tight-binding Hamiltonian on a square lattice constant $a$, and we set $e=\hbar=a=1$:

\begin{widetext}
\bea
H&=&\sum_{i,j\epsilon NN,\sigma} (-t+(-1)^{i_x+j_y} i\Delta_\mathrm{sf}(\vec r_{ij}))e^{iA_{ij}}c^\dag_{i\sigma}c_{j\sigma}+\sum_{i,j\epsilon NNN,\sigma}(-t')e^{iA_{ij}}c^\dag_{i\sigma}c_{j\sigma}+\nonumber\\
&&(\sum_{i,j\epsilon NN}(-1)^{i_x+j_x}\Delta_\mathrm{d}(\vec r_{ij})c^\dag_{i+} c^\dag_{j-}+H.C.)+
\sum_{i,\sigma}((-1)^{i_x+i_y}\sigma V_s(\vec r_i)+V_c(\vec r_i)-\mu)c^\dag_{i\sigma}c_{i\sigma},
\eea
\end{widetext}
where $t$ and $t'$ are the hoppings, $\Delta_\mathrm d$ is the nearest neighbor (NN) pairing, $\Delta_\mathrm{sf}$ is related to the staggered flux or equivalently d-density wave order (SF-DDW) \cite{thsu,Chakravarty01}, and $V_s$, $V_c$ are the staggered spin and charge potential respectively. The potentials are defined on sites $\vec{r_i}=(i_x,i_y)$ and the pairing is defined on bonds $\vec{r}_{ij}=(\vec{r}_i+\vec{r}_j)/2$. The $k$-dependent d-wave gap would be of size $2\Delta_\mathrm{d}(\cos k_x-\cos k_y)$, and near half filling the gap opening seen in the tunneling expriments would be $\sim 4\Delta_\mathrm{d}.$ As a phenomenological model, we take $t$ to be of order $J$, the antiferromagnetic coupling. In underdoped cuprates the gap size is then $\sim0.5t$. $A_{ij}$ is the electromagnetic gauge field taken to be compact on the lattice, which satisfies $\sum_{plaqutte}A_{ij}=B$ and $\sum_{triangle}A_{ij}=B/2$; $B$ is the magnetic field. 

The pairing amplitude near a vortex is described by the ansatz
\bea
\label{sinansatz}
|\Delta_\mathrm{d}(\vec r)|&=&\Delta_\mathrm{0}\sin\theta;\nonumber\\
\cos\theta(\vec r)&=&\frac{r_0}{\sqrt{r_0^2+d(\vec r)^2}},
\eea 
where $\Delta_0$ is the NN pairing amplitude deep inside the superconductor, $r_0$ the core size, and $d(\vec r)$ is the distance to the vortex center. When multiple vortices are near by we replace $d$ by $d_{min}=(\sum_id_i^{(-p)})^{-(1/p)}$ which is smaller than the distance toward the nearest vortex. The choice of $p\geq 1$ does not qualitatively affect the result. The phase $\phi$ of $\Delta_\mathrm{d}$ is not a gauge invariant quantity so we must determine it according to the gauge choice of $A_{ij}$. An alternative way is to assign $\phi$ and then determine the gauge field under the constraint mentioned above. We choose $\phi$ to follow the constraints: (i) $|\Delta\phi|<\pi/2$ for every link; (ii) $\sum_{loop}\Delta\phi=2\pi n$ where $n$ is the number of vortices enclosed, and (iii) $\phi$ is periodic in the $y$ direction. We then determine $A_{ij}$ using the fact that the physical configuration must minimize the free energy $\sum\frac12\Delta_\mathrm{d}^2(\vec r)v_s(!
 \vec r)^2$ where $v_s=\phi_i-\phi_j-A_{ij}$ is the superfluid velocity. The free energy is a function quadratic in $A_{ij}$, so we can optimize it by solving a linear equation, using sparse matrix routines. 

$V_c$, $\Delta_\mathrm{sf}$, and $V_s$ describe the order in the normal state. They can be uniform, periodic, or localized around
the vortices depending on the order present. There is an additional piece in $V_c$ which balances the charge density in the mixed state. In a finite system with periodic boundary conditions it is not always possible to fit in a periodic vortex lattice. Instead we work with with a disordered array of vortices. At each $B$ the vortex positions are determined by a Monte Carlo annealing process for particles with $Cr^{-2}$ repulsion. The vortices are stuck as we gradually lower the temperature. The resulting vortex configuration has short range order but is rather disordered. It is a reasonable representation of a snapshot of a vortex liquid or a pinned vortex solid.

In order to see the quantum oscillations, either the sample size has to be larger than the cyclotron radius, or one can take advantage of the periodic boundary conditions. With the periodic boundary conditions the magnetic flux through the sample is quantized, and to achieve a small stepping in $1/B$ the sample size again has to so large that the hamiltonian cannot be diagonalized. We can instead, use an iterative green's function's method to get the local density states (LDOS), at any fixed energy:

We affix our sample to two semi-infinite stripes in the $\pm x$ direction. The stripes are normal metal described by $t$ and $t'$. We take periodic boundary conditions in the $y$ direction. Now the configuration is similar to \cite{PALee81} and we can use the same method:
\bea
G^L(x)=&[G^0(x)^{-1}-\mathbf{t}G^L(x-1)\mathbf{t}^\dag]^{-1};\,\,\,\,\,\,\,\,\,\,\,\,\,\,
\,\,\,\,\,\,\,\,\,\,\,\,\,\,&\\
G(x)=&&\nonumber\\
\label{greens}
(G^0(x)^{-1}&-\mathbf{t}G^L(x-1)\mathbf{t}^\dag-\mathbf{t}^\dag G^R(x+1)\mathbf{t})^{-1},\,\,\,\,\,\,\,\,\,\,\,\,
\eea
where $G^0(x)$ is the Green's function for the isolated $x_{th}$ column, $G^L(x)$ ($G^R(x)$) is the Green's function at column $x$ when the right (left) side of the column is deleted, $\mathbf{t}$ is the hopping matrix between the two consecutive columns (contains $t, t', \Delta_\mathrm{d}(x)$, and $\Delta_\mathrm{sf}(x)$), and $G(x)$ is the Green's function for the $x_{th}$ column in the original setting. Everything here is a $2y\times2y$ matrix, containing the electron part and the hole part. We first go from left to right, compute $G^L$ for every $x$. Next we go from right to left to compute $G^R$ and use Eq (\ref{greens}) to compute $G(x)$ for all $x$. The imaginary part of the $y$th diagonal matrix element in the electron part of $G(x)$ is then related to the local density of states at $(x,y)$. 

Using this method we can get the LDOS everywhere on our sample. However, the LDOS varies from place to place due to its relative position to the vortices as well as the disorder of the vortex lattice. To see the quantum oscillations, it suffices to look at the averaged density of states. In this work, if not specified otherwise we set $t'=-0.3t$, $r_0=5$ and the lattice is of size $2000\times80.$ We vary the magnetic field in accord with the number of the vortices, which is from $500$ to $10000$.

\begin{figure}[!htp]
 		\includegraphics[scale=0.32]{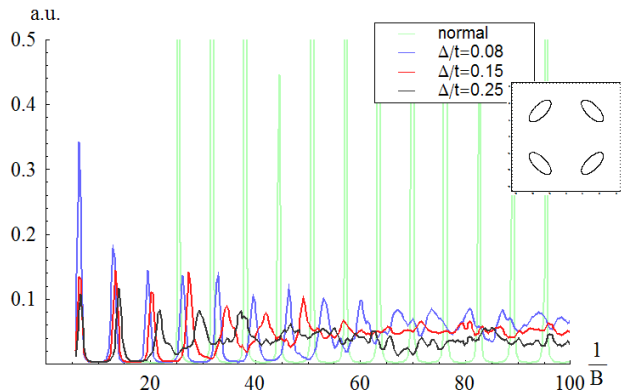}
		\caption{\label{sfhp}this plot shows the averaged DOS vs. $1/B$ in the vortex mixed state with $(\pi,\pi)$ SF-DDW order. $B$ is in units of $(\frac{\Phi_0}{2\pi a^2})$.  If we view $H_{c2}$ as the field strength where vortices are overlapping, it corresponds to $1/B=25$ in the plot. Different lines in the plot show states with different pairing fields. The maximum pairing gap in the antinodal direction would have size $\sim 4\Delta.$ We can see as the gap increases the frequency decreases. The inset is a plot of the Fermi surface in the normal state. There are 4 hole pockets with 2.5\% area each of the original BZ. This corresponds to $p=0.1$ from half filling. In this plot $\Delta_\mathrm{sf}=0.25t$.}
 \end{figure}

In Fig. \ref{sfhp} we showed the result where we start from the state in which the Fermi surface is reconstructed by $(\pi,\pi)$ SF-DDW order. If we define $H_{c2}$ to be the field where vortices start overlapping and $r_0=5a$, this gives $(\frac{\Phi_0}{2\pi a^2})\frac1{H_{c2}}=25$, where $\Phi_0$ here is the full flux quantum, $hc/e$. The experimental probe, roughly at the $10$th Laudau level, would be around $(\frac{\Phi_0}{2\pi a^2})\frac1{H_{c2}}=60.$ We create a gap large enough to kill the electron pockets, leaving 4 hole pockets shown in the inset. We found that the oscillation period of the normal state matches the prediction from the Onsager relation and the Luttinger theorem. As we turn on the d-wave pairing amplitude, the period of the quantum oscillation increases. As the pairing gets large the period of the oscillation is off from the period in the normal state by about 20\%. The frequency shifts are clearly seen in the Fourier transform shown in Fig. \ref{fourier}. Note that in our modelling of the vortex cores, the superconducting pairing remains, albeit at reduced amplitude, even for  $B>H_{c2}.$ This explains why the frequency shift persists somewhat above $H_{c2}$, but we can ignore that region becouse it is not reached experimentally. We have also done calculations for a $(\pi,\pi)$ spin density wave (SDW) order, and the period is shifted in a similar way (see Fig. \ref{sdwhf}.)

\begin{figure}
 	\includegraphics[scale=0.32]{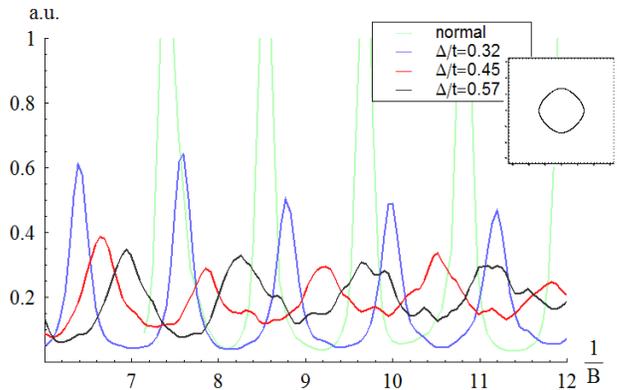}
	\caption{\label{fig:ep0} an electron pocket is centered at the origin. The pocket has an area of 14\% of the original BZ. In this plot $t'=-0.14t.$ The maximum gap $\sim 1.5\Delta.$}
\end{figure}

To check whether this is a generic phenomena, we ran a simpler setting with an electron pocket centered at origin. The result is showed in Fig. \ref{fig:ep0}. 
Again, as we turned on the d-wave pairing, the frequency is reduced as the pairing is increased. It is worth noting that we shifted the chemical potential in order to maintain the total electron density as we increase the pairing. We also added a charge potential to make the charge density approximately uniform inside and outside the vortex core. Even if we just forget about density and keep the same chemical potential as in the normal state, the period still decreases (not shown). The chemical potential shift is significant in the case of Fig. \ref{fig:ep0} because the pairing is strong and the electron is far from particle-hole symmtries. It is less significant in the more realistic case of Fig. \ref{sfhp}.
 
\begin{figure}[htp]
 	\includegraphics[scale=0.32]{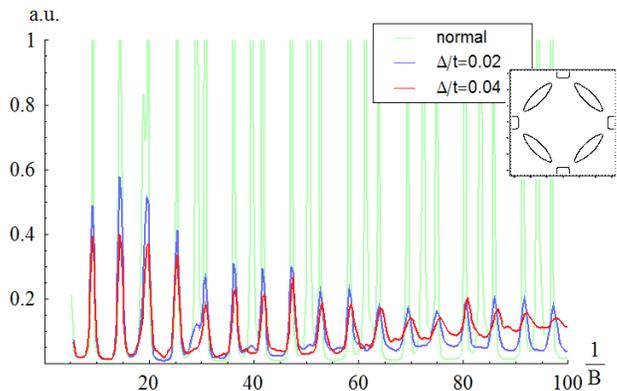}
\caption{\label{sdwhpep}\,4 hole pockets and 2 electron pockets are present. In this plot $A_h=2.9\%$ and $A_e=1.44\%$ of the original BZ, and $p=0.086$. In the normal state, from Onsager's relation we can determine that the peaks with period $11$ are from electron pockets. The amplitude is heavily suppressed in the mixed state. In this plot $V_s=0.2t$.}
\end{figure}

In Fig. \ref{sdwhpep} we started with a 2-pocket Fermi surface reconstructed by a $(\pi,\pi)$ SDW order, and turned on a small d-wave pairing. The oscillations from the electron pockets are clearly visible in the normal state, but are rapidly killed in the mixed state. Note the pair field is very small, about 10 times smaller than that in Fig. \ref{sfhp}. One explanation is that the electron pocket originates from the antinodal region where the gap is large, and is dephased by the random pairing potential due to the random vortex configurations. However, this cannot be the whole story because we find that for s-wave pairing with similar pairing gap size the electron and hole pockets both survive. One difference we noticed is that in the s-wave case there is a strong density of states peak inside the vortex core, much larger than that inside the d-wave core. At this point we do not have a full understanding of the suppression of the electron pocket oscillations.  

We have also calculated the oscillation for an "incommensurate" SDW, with period $Q=(\pi(1\pm 2\delta),\pi)$ where $\delta=1/8.$ We impose a sinusoidal $V_s(\vec{r})=V_{s0}\cos(2\pi\delta x))$ and $V_c(\vec{r})=V_{c0}\cos(4\delta x)$. Complicated band structures in this kind of potential was computed by Millis and Norman \cite{Millis07}. It is instructive to consider the hybridization only with the primary vector $Q$ for SDW and $2Q$ for the charge component.  \begin{figure}[!h]
\includegraphics[scale=0.3]{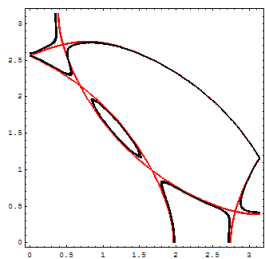}
\caption{\label{bs}Primary bands for the incommensurate SDW in the first quadrant of the original BZ.}
\end{figure}
Fig. \ref{bs} shows the pure SDW case ($V_c=0$) and we see that after hybridization only two closed orbits are possible. One is a small hole pocket and the other is the electron pocket. The larger hole pocket becomes an open orbit. Inclusion of $V_c$ further cuts the small hole pocket to an even smaller area. 
\begin{figure}[!h]
\includegraphics[scale=0.32]{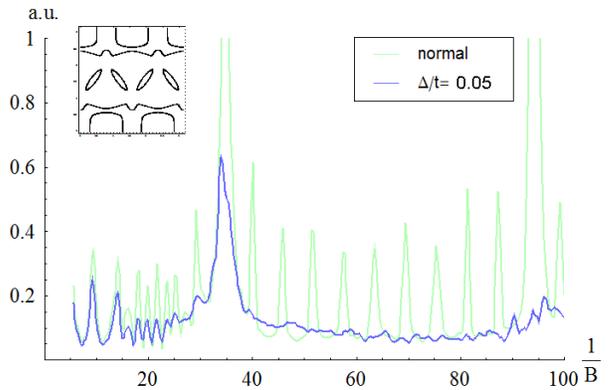}
\caption{\label{isdw} 1/8 stripe state with $t'=-0.4t$, $V_{s0}=0.4$, and $V_{c0}=0.1$. In the normal state 3 frequencies are visible. The fast oscillations localized around $1/B\sim 20$ is a breakdown effect. At lower fields the fast oscillation (period 6) is coming from the electron pockets and the very slow oscillation (only 1 period seen in the range) is from the hole pockets. As we turn on superconductivity the oscillation from the electrons is diminished.}
\end{figure}
Fig. \ref{isdw} shows the quantum oscillations in a uniform $1/8$ stripe phase. We indeed see the oscillations from the electron pockets and the small hole pockets. The electron pocket is again heavily suppressed in the mixed state, leaving a hole frequency which is much too small compared with experiment.

\begin{figure}[htp]
	
	\subfigure[\label{sff}]{
		
 		\includegraphics[scale=0.25]{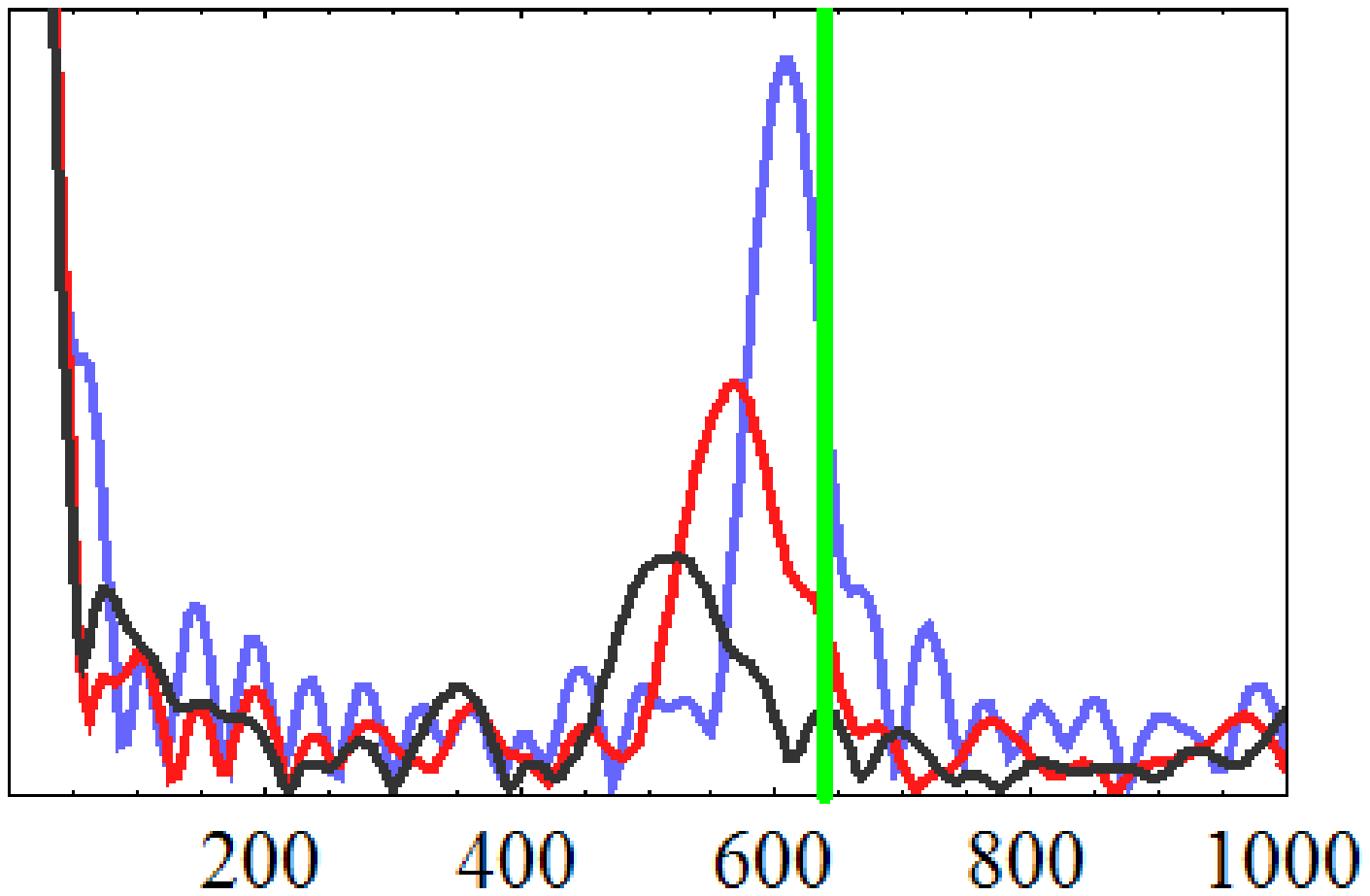}
	}	
\subfigure[\label{sdwhf}]{
		
 		\includegraphics[scale=0.25]{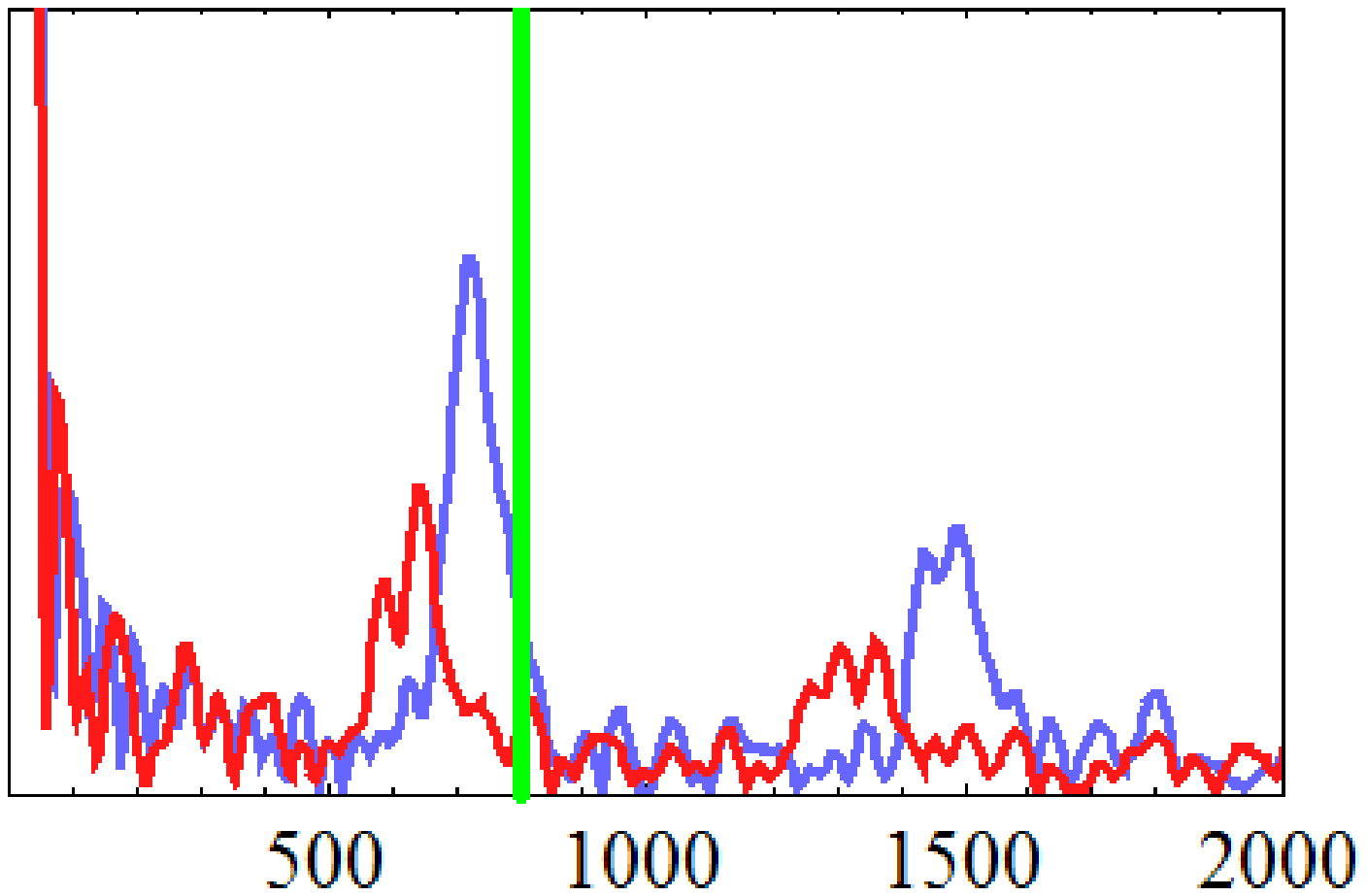}
	}
\subfigure[\label{epf}]{
		
 		\includegraphics[scale=0.25]{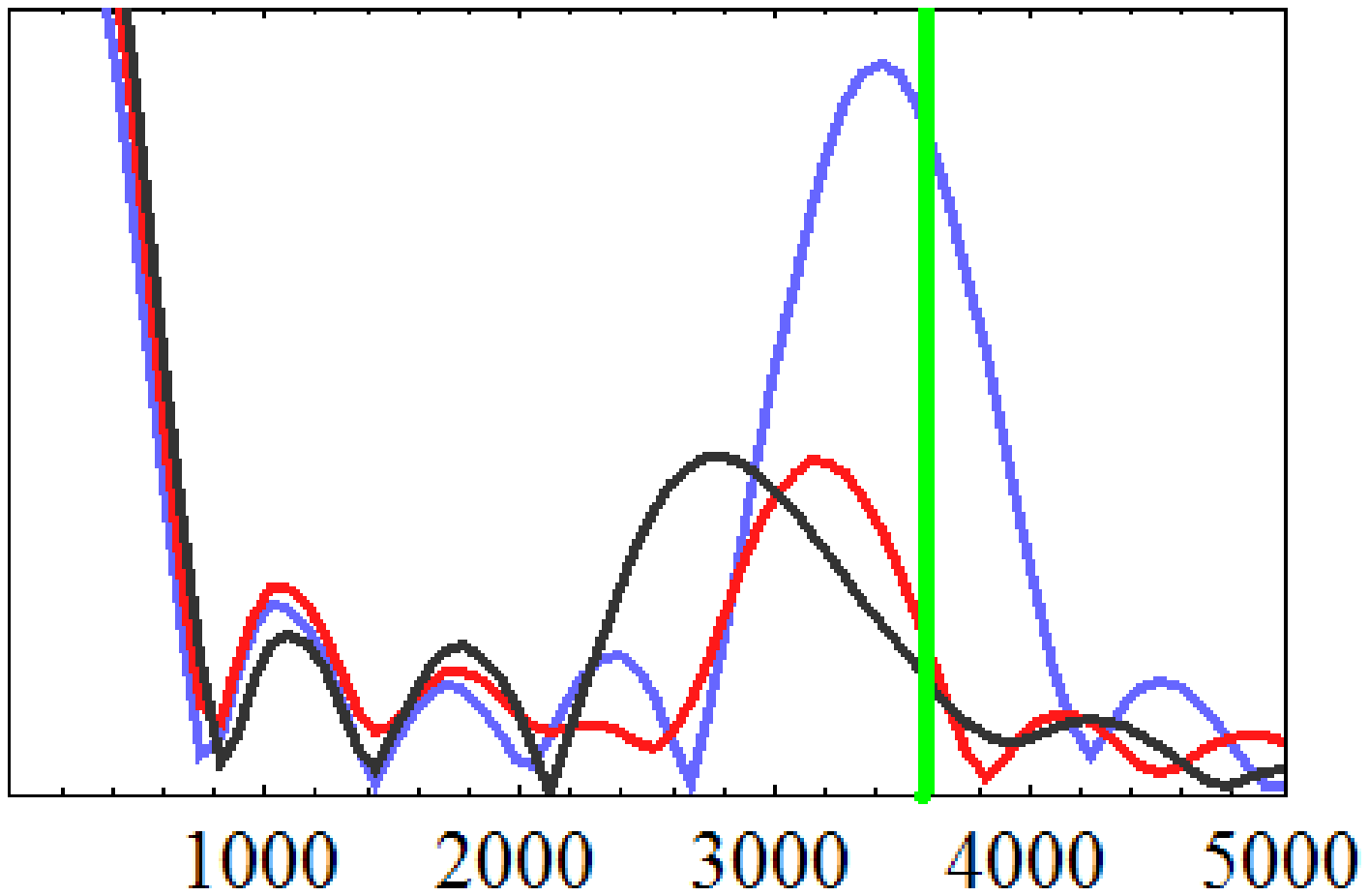}
	}
	\subfigure[\label{sdwhef}]{
		
 		\includegraphics[scale=0.25]{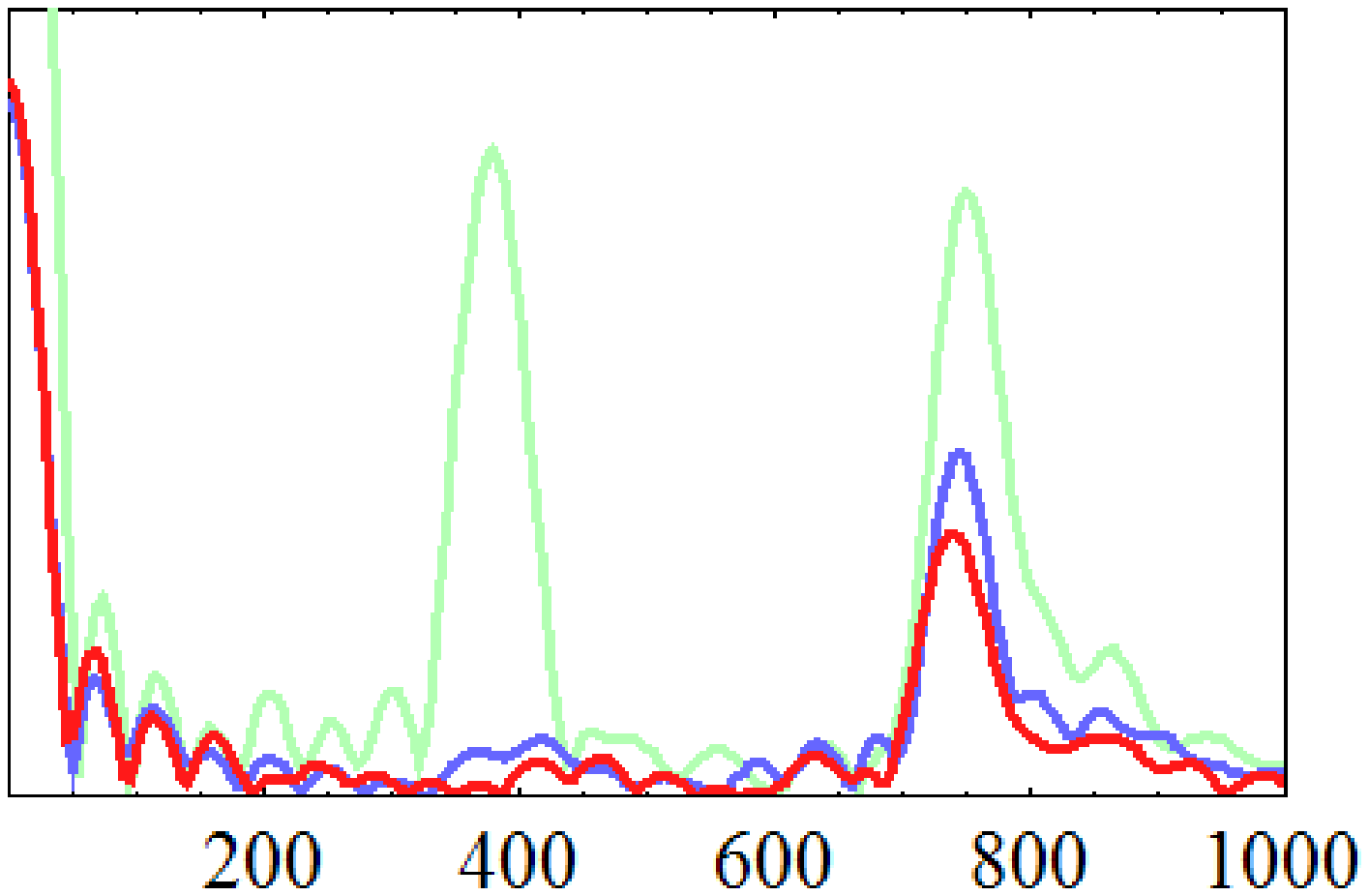}
	}	
 \caption{\label{fourier} Fourier transforms of  plots shown in Fig \ref{sfhp}, \ref{fig:ep0}, and \ref{sdwhpep}. We take $a=4\,A^\circ$ and the frequency is then in units of Tesla. The color used is the same as in the averaged DOS vs. 1/B plots. The normal state frequency is shown by the green line. (a) SF-DDW.(b) SDW, p=0.125. The peak at the right is the second harmonic. (c) electron pocket centered at origin. (d) two-pocket SDW; the peak at $\sim$380 T is from the electron pockets and the one at $\sim$750T is from the hole pockets.}
 \end{figure}

In conclusion we have shown that the Onsager relation does not always apply to quantum oscillations in the vortex mixed state. When the pairing is strong and the coherence length short, we find a systematic decrease in the frequency. We have also checked that if the core size is increased, the shift is diminished. Thus our result is consistent with experiments performed so far on conventional superconductors, which are in the large core size limit. The implication for the experiment on high $T_c$ cuprate is that the simple interpretation in terms of $(\pi,\pi)$ folding creating four hole pockets cannot be automatically ruled out. There are two candidates for $(\pi,\pi)$ order. One is SDW order and the second is SF-DDW order. We note that in view of the recent measurement that the $g$ factor is much less than 2 \cite{sebastianpr}, the SF-DDW scenario must be accompanied by some additional order, such as incommensurate SDW. In that case there are possibilities for larger orbits composed of two or three hole pockets shifted by $\delta$ if magnetic breakdown is taken into account. The other interpretation is the incommensurate SDW order. In this case the observed pocket must be identified with the electron pocket. Alternatives in terms of SF-DDW with electron porckets have also been proposed. \cite{Chakravarty08,Jdimov,podolsky}. While the electron pocket is rapidly suppressed in the mixed state in our model, it is not clear how general this conclusion is, i.e. whether it is valid beyond the BdG theory. We think that within the Fermi liquid scenario both are viable options at this point and further work will be needed to distinguish between them.

\begin{acknowledgments}
We thank T. Senthil for many helpful discussions. This work is supported by NSF Grant No. DMR08-04040.
\end{acknowledgments}


\end{document}